\documentstyle[prl,aps,preprint]{revtex}
\newcommand{\be}{\begin{eqnarray}}
\newcommand{\ee}{\end{eqnarray}}

\begin{document}
\draft
\title{Dimensional Crossover driven by Magnetic Ordering in Optical Conductivity of
Pr$_{1/2}$Sr$_{1/2}$MnO$_3$}
\author{J. H. Jung,$^1$ E. J. Choi,$^2$ H. J. Lee,$^1$ Jaejun Yu,$^1$ Y. Moritomo,$%
^3 $ and T. W. Noh$^1$}
\address{$^1$Department of Physics, Seoul National University, Seoul 151-742,\\
Korea}
\address{$^2$Department of Physics, University of Seoul, Seoul 130-743, Korea}
\address{$^3$CIRSE, Nagoya University, Nagoya 464-8603 and PRESTO, JST, Japan}
\maketitle

\begin{abstract}
We investigated optical properties of Pr$_{0.5}$Sr$_{0.5}$MnO$_3$, which has
the $A$-type antiferromagnetic ordering at a low temperature. We found that $%
T$-dependence of spectral weight transfer shows a clear correlation with the
magnetic phase transition. In comparison with the optical conductivity
results of Nd$_{0.5}$Sr$_{0.5}$MnO$_3$, which has the CE-type
antiferromagnetic charge ordering, we showed that optical properties of Pr$%
_{0.5}$Sr$_{0.5}$MnO$_3$ near the N\'{e}el temperature could be explained by
a crossover from 3D to 2D metals. Details of spectral weight changes are
consistent with the polaron picture.
\end{abstract}





\newpage 

Optical properties of colossal magnetoresistance manganites have been
investigated extensively, since they showed drastic spectral weight changes
with temperature variation\cite{okimoto,kaplan,quijada,kim1,jung98,kim2}.
Especially, in ferromagnetic (FM) metallic states, their optical
conductivity spectra $\sigma (\omega )$ showed a large incoherent absorption
peak near 0.5 eV. It has been argued that the anomalous incoherent
absorption should be related to the Jahn-Teller polaron\cite
{kaplan,quijada,kim1,jung98,kim2,millis}. However, there are alternative
scenarios which attribute the incoherent feature to orbital degree of freedom%
\cite{okimoto,ishihara,shiba,kilian}. Recently, Mack and Horsch investigated
a planar 2D model for FM phase of manganites which develops orbital order of 
{\it e}$_g$ electrons with $x^2-y^2$ symmetry\cite{mack99}. With the
finite-temperature diagonalization method, they calculated dynamic structure
factor of orbital excitations and $\sigma (\omega )$, and suggested that $%
\sigma (\omega )$ in the doped $x^2-y^2$ ordered phase can allow one to
distinguish between the orbital excitation and the Jahn-Teller polaron
scenarios.

A recent neutron scattering study on Pr$_{0.5}$Sr$_{0.5}$MnO$_3$ (PSMO)
provided a surprising result that the heavily doped 3D manganite has a
ground state with an antiferromagnetic (AF) {\it A}-structure\cite{kawano}.
In this state, FM planes are coupled antiferromagnetically, and conduction
should occur within the planes. And, the occupied orbitals were suggested to
have $x^2-y^2$ symmetry, so optical investigation could provide an answer to
the test proposed by Mack and Horsch. Moreover, the compound has two
interesting phase transitions: an AF-to-FM transition of the first order at $%
T_N\approx 150$ K, and a FM metal-to-paramagnetic insulator transition at $%
T_C\approx 270$ K\cite{tomioka}.

Another half-doped Nd$_{0.5}$Sr$_{0.5}$MnO$_3$ (NSMO) compound was found to
exhibit similar phase transitions. As $T$ increases, NSMO experiences an AF
insulator-to-FM metal transition at $T_N\approx 160$ K and a FM
metal-to-paramagnetic insulator transition at $T_C\approx 250$ K. In fact,
the low temperature AF insulator phase of NSMO becomes metallic under a high
magnetic field\cite{kuwahara}, and a similar transition was observed in PSMO%
\cite{tomioka}. Despite these similar features, the low temperature phase of
NSMO accompanies the {\it CE}-type AF ordering with charge and orbital order%
\cite{kawano}.

With the {\it CE}-type ordering, small hopping energy of holes along the 1D
FM chain is dominated by the charge- or orbital-interaction (also combined
with electron-phonon interaction), leading to a charge ordering. On the
contrary, holes in the 2D FM layer of PSMO with the {\it A}-type ordering
may have larger hopping energies, possibly leading to a 2D metallic state
competing with the charge-ordering instability. Since the conduction in the
FM metallic region is 3D, a dimensional crossover could occur at $T_N$.
However, such a dimensional crossover due to the magnetic ordering has not
been yet investigated in PSMO.

In this letter, we report the first results of the temperature dependent $%
\sigma (\omega )$ for PSMO. The variation of $\sigma (\omega )$ across $T_N$
in PSMO are compared with that in NSMO, and it can be well understood in
terms of a dimensional crossover driven by the {\it A}-type AF ordering from
a 3D FM metallic state above $T_N$ into a 2D metallic state below $T_N$.
Details of spectral weight changes are consistent with the polaron picture.

Single crystals of PSMO and NSMO were grown by the floating zone method\cite
{moritomo}. For optical measurements, both samples were polished up to 0.3 $%
\mu m$ using diamond pastes. To release strain applied during the polishing
procedure, they were annealed in an oxygen atmosphere\cite{haeja}. Near
normal incident reflectivity spectra $R(\omega )$ were measured from 8 meV
to 6.0 eV at the temperature region between 15 K and 300 K, and $\sigma
(\omega )$ were obtained using the Kramers-Kronig transformation. The
results were also confirmed independently by spectroscopic ellipsometry
techniques. Details of the optical measurements were published previously%
\cite{jung97}.

Figure 1(a) shows $\sigma (\omega )$ of PSMO in the mid-infrared region. The
sharp peaks below 0.1 eV represent optic phonon modes. Above $T_C$, there is
a broad peak-like feature around 0.8 eV, followed by a rise above 2.0 eV.
With cooling into the FM state, the spectral weight above 0.6 eV is
transferred to a lower energy region. With further cooling below $T_N$, the
spectral weight changes move in the opposite direction, i.e. from lower to
higher energy regions.

Further insights on the spectral weight changes can be obtained from $\Delta
\sigma (\omega )\equiv \sigma (\omega ,T)-\sigma (\omega ,T=290K)$, shown in
Fig.1(b). Overall spectral weight changes in $\Delta \sigma (\omega )$ occur
around 0.4 eV and 1.5 eV, which suggests that $\sigma (\omega )$ could be
interpreted in terms of two-peak structures\cite{jung98,dagotto}. As $T$
decreases, $\Delta \sigma (\omega )$ around 0.4 eV increases and reaches a
maximum near $T_N$. Then, it decreases and remains finite at 15 K. As shown
in the inset, a similar $T$-dependence was observed in $\Delta \sigma
(\omega )$ of NSMO. However, $\Delta \sigma (\omega )$ around 0.4 eV becomes
nearly zero at 15 K, which is different from the PSMO case.

Figure 2 displays $\sigma (\omega )$ of PSMO in the far-infrared (FIR)
region. [For clarity, the $\sigma (\omega )$ curve at 290 K only were
plotted by shifting upward by 800 $\Omega ^{-1}$cm$^{-1}$.] The peaks around
160 cm$^{-1}$, 330 cm$^{-1}$, and 580 cm$^{-1}$ represent infrared-active
phonons which correspond to the external, bending, and stretching modes,
respectively\cite{kim96}. Around $T_C$, there are little changes in their
characteristics. However, below $T_N$, the bending and the stretching modes
are shifted by about 5 cm$^{-1}$ and 8 cm$^{-1}$, respectively, and their
strength becomes increased. The phonon frequency changes agree with
significant changes in lattice parameters accompanying the AF transition\cite
{kawano}. Below $200$ cm$^{-1}$, Drude-like peaks in $\sigma (\omega )$ do
appear for all temperatures. Even at 15 K, the Drude-like feature still can
be seen.

The far-infrared $\sigma (\omega )$ of NSMO, displayed in the inset of Fig.
2, are different from those of PSMO. First, although the room temperature
phonon spectrum of NSMO is similar to that of PSMO, the bending and the
stretching phonon peaks are split strongly below $T_N$. These phonon
splittings are very similar to those of other manganites in the {\it CE}%
-type charge ordered state\cite{ishikawa}. Second, the Drude peak of NSMO
vanishes completely at 15 K. The absence of the Drude peak is consistent
with its high {\it dc} resistivity value (i.e. about 0.43 $\Omega cm$),
which arises from a strong carrier localization due to the charge ordering.
The absence of the phonon splittings and the presence of the finite Drude
component at $T\ll T_N$ in PSMO suggests that the electronic ground states
are quite different in these two systems.

To get a better understanding on the $T$-dependent spectral weight changes,
we decomposed the electronic contributions in $\sigma (\omega )$ into three
parts: the FIR Drude component $\sigma _{Drude}(\omega )$, the midgap
component $\sigma _{ms}(\omega )$, and the charge transfer component $\sigma
_{CT}(\omega )$. 
\begin{equation}
\sigma (\omega )=\sigma _{Drude}(\omega )+\sigma _{ms}(\omega )+\sigma
_{CT}(\omega ).
\end{equation}
The gradual rise at $\omega >2$ eV in Fig.1(a) corresponds to an onset of $%
\sigma _{CT}(\omega )$, and we used the Lorentz oscillator model to fit this
energy region. Since the significant changes of spectral weight occur around
0.4 eV and 1.5 eV, as shown in Fig.~1(b), we used two Gaussian peaks for
fitting. We set the peak positions to be 0.5 eV (for Peak I) and 1.5 eV (for
Peak II). Note that similar analyses were performed on La$_{1-x}$Ca$_x$MnO$_3
$ quite successfully\cite{jung98}.

Figure 3 shows the experimental data and the fitting results of $\sigma
(\omega )-\sigma _{CT}(\omega )-\sigma _{Drude}(\omega )$. Agreement between
the experimental and the fitting results is quite good. A strength of each
peak was obtained by integrating the corresponding spectral weight. The
solid circles in Fig. 4(a) and (b) denote S$_I$ and S$_{II}$, which are
strengths of Peak I and Peak II, respectively. As $T$ decreases, S$_I$
increases slightly around $T_C$ and then begins to decrease near $T_N$. The $%
T$-dependence of S$_{II}$ is roughly the opposite. Fig. 4(c) shows the $T$%
-dependence of Drude peak strength, S$_D$, which is similar to that of S$_I$%
. For comparisons, peak strengths for NSMO were also plotted with solid
squares in Fig. 4(a), (b), and (c). In NSMO, the $T$-dependences of S$_I$, S$%
_{II}$, and S$_D$ are much stronger and S$_D$ becomes nearly zero below $T_N$%
.

Mack and Horsch investigated a planar model for the FM phase which develops
orbital order of {\it e}$_g$ electrons with $x^2-y^2$ symmetry\cite{mack99}.
Based on the orbital{\it \ t}-{\it J} model, they provided a theoretical
prediction for $\sigma (\omega )$ in the 2D FM state: $\sigma (\omega )$
should show both a Drude peak and a gapped incoherent absorption. The
incoherent absorption due to the orbital excitation is expected to be peaked
at 2$t$, where $t$ is the hopping energy. With $t\sim $ 0.25 eV, Peak I
should correspond to the incoherent absorption, and the decrease of S$_I$
below $T_N$ is consistent with their prediction. However, for the coherent
peak, S$_D$ is predicted to increase as $T$ is lowered, opposite to our
observation in Fig.4(c). Therefore, our data cannot be explained in terms of
the orbital excitation only.

Another interesting scenario is the Jahn-Teller polaron picture\cite
{quijada,kim1,jung98,millis}, where the Drude peak and Peak I can be
interpreted as coherent and incoherent absorption peaks due to polaron
motion, respectively. [Namely, Peak I can be assigned as the inter-atomic 
{\it e}$_g^1$(Mn$^{3+}$) $\rightarrow $ {\it e}$_g$(Mn$^{4+}$) transition,
where {\it e}$_g^1$ represents the lower energy orbital of the Jahn-Teller
split levels.] The increase of S$_I$ at $T<T_C$ can be interpreted in terms
of the FM alignment of the Mn moments, since number of available hopping
sites will increase. In the FM state, there will be a crossover from small
to large polaron states\cite{kim1}. Coherent motion of carriers and far-IR
Drude absorption will increase, which is in good agreement with Fig. 4(c).
Below $T_N$, the AF ordering will prevent polaron hopping between
neighboring Mn sites with opposite spins, so S$_I$ and S$_{Drude}$ should
decrease, consistent with Fig. 4(a) and (c).

In earlier works\cite{kaplan,jung98}, Peak II was assigned as a transition
between the Jahn-Teller split levels: i.e. {\it e}$_g^1$(Mn$^{3+}$) $%
\rightarrow $ {\it e}$_g^2$(Mn$^{3+}$). Assuming that this transition is
intra-atomic\cite{note1}, the temperature dependence of S$_{II}$ can be
explained. The intra-atomic transition becomes possible due to the Mn {\it e}%
$_g-$O 2{\it p } hybridization and the local distortion of the MnO$_6$
octahedron. In the metallic states, the Jahn-Teller distortion becomes
weakened due to increased screening, so the MnO$_6$ octahedron becomes more
close to cubic and S$_{II}$ will be decreased. Below $T_N$, the local
symmetry becomes lowered again\cite{kawano}, and S$_{II}$ becomes enhanced,
consistent with Fig.4(b).

Note that the $T$-dependences of S$_I$, S$_{Drude}$ and S$_{II}$ of PSMO are
weaker than those of NSMO. This implies that metallic screening is more
effective, particularly for $T<T_N$. The prevalence of the screening effects
indicate that in the {\it A}-type AF phase of PSMO, the FM layers can form
2D metallic layers with c-axis conduction forbidden. The existence of the
Drude-like feature at 15 K, displayed in Fig. 2, and the finite value of S$%
_{Drude}$ below $T_N$, displayed in Fig. 4(c), also supports that PSMO
should remain in a metallic state with the AF ordering. Then, the
resistivity change of PSMO at $T_N$ should be interpreted as a dimensional
crossover from 3D to 2D metallic states driven by the {\it A}-type AF
ordering.

To further investigate this intriguing phenomenon, we looked into total
absorption strength due to the polaron motion, i.e., S$_{tot}=$ S$_{Drude}$
+ S$_I$, which is shown in Fig. 4(d). As $T$ decreases, S$_{tot}$ starts to
increase around $T_C$ and then begins to decrease at $T_N$. In (La,Pr)$_{0.7}
$Ca$_{0.3}$MnO$_3$, whose ground state remains a 3D FM metal at a low $T$,
it was found that S$_{tot}$($T$) could be scaled with the double exchange
bandwidth $\gamma _{DE}$($T$)$=\langle \cos (\theta _{ij}/2)\rangle $\cite
{kim2}, where $\theta _{ij}$ is the relative angle of neighboring spins and $%
\langle $ $\rangle $ represents thermal average in the double exchange model%
\cite{kubo}. This scaling behavior was explained in a model by R\"{o}der 
{\it et al.}\cite{roder}, where the double exchange and the Jahn-Teller
polaron Hamiltonians were taken into account. The dotted line in Fig. 4(d)
shows $\gamma _{DE}$($T$) for the 3D metal. Above $T_N$, agreement between S$%
_{tot}$($T$) and $\gamma _{DE}(T)$ is quite good.

However, S$_{tot}$($T$) deviates from $\gamma _{DE}$($T$) below $T_N$. This
deviation can be explained by the dimensional crossover from 3D to 2D due to
the AF ordering. In the 3D FM state, around a Mn ion, there are six
neighboring Mn sites with parallel spins, so they are all available for
polaron hopping. In the 2D AF state, four of them have parallel spins, but
the rest two have opposite spins. If we neglect the dependence of hopping on
orbital directions, S$_{tot}$ in 2D should be 2/3 of S$_{tot}$ in 3D. [Note
that, due to the multi-domain nature of the crystal, we are not probing the
in-plane and the c-axis conductivities separately but a statistical average
of the two.] The ratio between the experimental value of S$_{tot}$(15 K) and
its theoretical value predicted from $\gamma _{DE}$(15 K) is found to be
(0.72$\pm $0.06), which is close to 2/3=0.67. The small deviation might come
from the hopping dependence on orbital directions and/or contribution of the
orbital excitation.

We want to address some interesting points for future studies. First, S$%
_{Drude}$ is much smaller than S$_I$. Second, the $T$-dependence of S$%
_{Drude}$ is stronger than that of S$_{tot}$($T$) [or S$_I$]. As $T$
decreases from $T_N$ to 15 K, S$_{Drude}$ decreases by a factor of about 3.
A detailed model which includes both the orbital excitation and the polaron
is highly desirable to get further understanding on $\sigma (\omega )$. A
recent calculation by Kilian {\it et al.}\cite{kilian} suggested that $%
\sigma (\omega )$ of (La,Sr)MnO$_3$ could be explained better by including
the orbital and the lattice degrees of freedom together. Finally, it would
be highly desirable to investigate the anisotropic nature of the 2D metallic
state using single domain PSMO samples. Similar studies were done very
recently on Nd$_{0.45}$Sr$_{0.55}$MnO$_3$\cite{kuwahara99}.

In summary, we investigated optical properties of Pr$_{0.5}$Sr$_{0.5}$MnO$_3$%
, which has the {\it A}-type antiferromagnetic ordering at a low
temperature. Its temperature dependent spectral weight change shows a clear
correlation with the magnetic phase transition. Its spectral weight changes
below the N\'{e}el temperature could be explained by a crossover driven by
the magnetic ordering from 3D to 2D metal.

We acknowledge Professor J.-G. Park for discussion. This work was supported
by Ministry of Science and Technology through grant No. I-3-061, by Seoul
National University Research Fund, and by the Korea Science \& Engineering
Foundation through RCDAMP of Pusan National University. The work by Y.M. was
supported by a Grant-In-Aid for Scientific Research from the Ministry of
Education, Science, Sports and Culture, and from PRESTO, JST.

\begin{figure}[tbp]
\caption{(a) $T$-dependent optical conductivity $\sigma (\omega )$, and (b) $%
\Delta \sigma (\omega )\equiv $ [$\sigma (\omega ,T)-\sigma (\omega ,290K)$]
of PSMO. In the inset, $\sigma (\omega )$ of NSMO are shown. }
\label{Fig:2}
\end{figure}

\begin{figure}[tbp]
\caption{$T$-dependent $\sigma (\omega )$ of PSMO in the far-infrared
region. For clarity, $\sigma (\omega )$ at 290 K are plotted by moving
upward by 800 $\Omega ^{-1}$cm$^{-1}$. In the inset, $\sigma (\omega )$ of
NSMO are shown. The dotted lines represent the phonon frequencies at 290 K. }
\label{Fig:3}
\end{figure}

\begin{figure}[tbp]
\caption{Midgap state optical conductivity $\sigma _{ms}(\omega )$ of PSMO.
The solid circles, the dotted lines, and the solid lines are represent the
experimental data, the Gaussian functions, and the sums of two Gaussian
functions, respectively. }
\label{Fig:4}
\end{figure}

\begin{figure}[tbp]
\caption{Optical strengths of (a) S$_I$, (b) S$_{II}$, (c) S$_{Drude}$, and
(d) S$_I$+S$_{Drude}$. In (d), the dotted line represents the prediction
from the $T$-dependent double exchange bandwidth. The deviation of (S$_I$+S$%
_{Drude}$) near $T_N$ can be attributed to the dimensional crossover driven
by magnetic ordering.}
\label{Fig:5}
\end{figure}


\begin{references}
\bibitem{okimoto}  Y. Okimoto, {\it et al.}, Phys. Rev. Lett. {\bf 75}, 109
(1995); Phys. Rev. B {\bf 55}, 4206 (1997).

\bibitem{kaplan}  S. G. Kaplan, {\it et al.}, Phys. Rev. Lett. {\bf 77},
2081 (1996).

\bibitem{quijada}  M. Quijada, {\it et al.}, Phys. Rev. B {\bf 58}, 16 093
(1998).

\bibitem{kim1}  K. H. Kim, {\it et al.}, Phys. Rev. Lett. {\bf 81}, 1517
(1998).

\bibitem{jung98}  J. H. Jung, {\it et al.}, Phys. Rev. B {\bf 57}, R11 043
(1998).

\bibitem{kim2}  K. H. Kim, {\it et al.}, Phys. Rev. Lett. {\bf 81}, 4983
(1998).

\bibitem{millis}  A. J. Millis, {\it et al.}, Phys. Rev. B {\bf 54}, 5405
(1996).

\bibitem{ishihara}  S. Ishihara, {\it et al.}, Phys. Rev. B {\bf 56}, 686
(1997).

\bibitem{shiba}  H. Shiba, {\it et al.}, J. Phys. Soc. Jpn. {\bf 66}, 941
(1997).

\bibitem{kilian}  R. Kilian and G. Khaliullin, Phys. Rev. B 58, R11 841
(1998).

\bibitem{mack99}  F. Mack and P. Horsch, Phys. Rev. Lett. {\bf 82}, 3160
(1999).

\bibitem{kawano}  H. Kawano, {\it et al.}, Phys. Rev. Lett. {\bf 78}, 4253
(1997).

\bibitem{tomioka}  Y. Tomioka, {\it et al.}, Phys. Rev. Lett. {\bf 74}, 5108
(1995).

\bibitem{kuwahara}  H. Kuwahara, {\it et al.}, Science {\bf 270}, 961 (1995).

\bibitem{moritomo}  Y. Moritomo, {\it et al.}, Phys. Rev. B {\bf 55}, 7549
(1997).

\bibitem{haeja}  H. J. Lee, {\it et al.} (submitted Phys. Rev. B,
cond\_matt/9904173).

\bibitem{jung97}  J. H. Jung, {\it et al.}, Phys. Rev. B {\bf 55}, 15 489
(1997).

\bibitem{dagotto}  S. Yunoki, A. Moreo, and E. Dagotto, Phys. Rev. Lett. 
{\bf 81}, 5612 (1998).

\bibitem{kim96}  K. H. Kim, {\it et al.}, Phys. Rev. Lett. {\bf 77}, 1877
(1996).

\bibitem{ishikawa}  T. Ishikawa, et al. Phys. Rev. B {\bf 59}, 8367 (1999).

\bibitem{note1}  Some workers claimed that Peak II comes from inter-atomic
transition between the Jahn-Teller split levels. Refer to Ref. 3.

\bibitem{kubo}  K. Kubo and N. Ohata, J. Phys. Soc. Jpn. {\bf 33}, 21 (1972).

\bibitem{roder}  H. R\"{o}der, J. Zhang, and A. R. Bishop, Phys. Rev. Lett. 
{\bf 76}, 1356 (1996).

\bibitem{kuwahara99}  H. Kuwahara, {\it et al.}, Phys. Rev. Lett. {\bf 82},
4316 (1999).
\end{references}
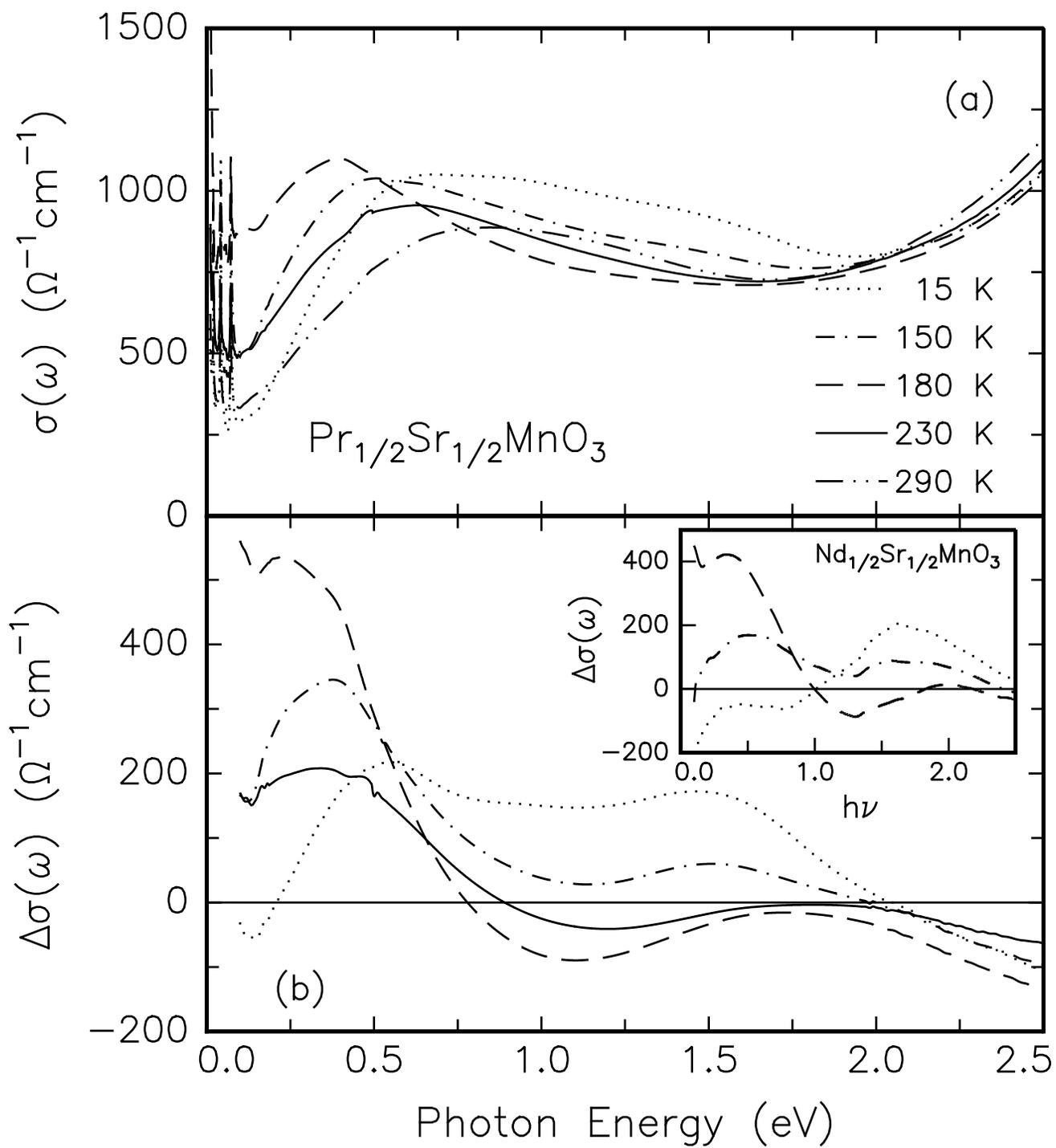
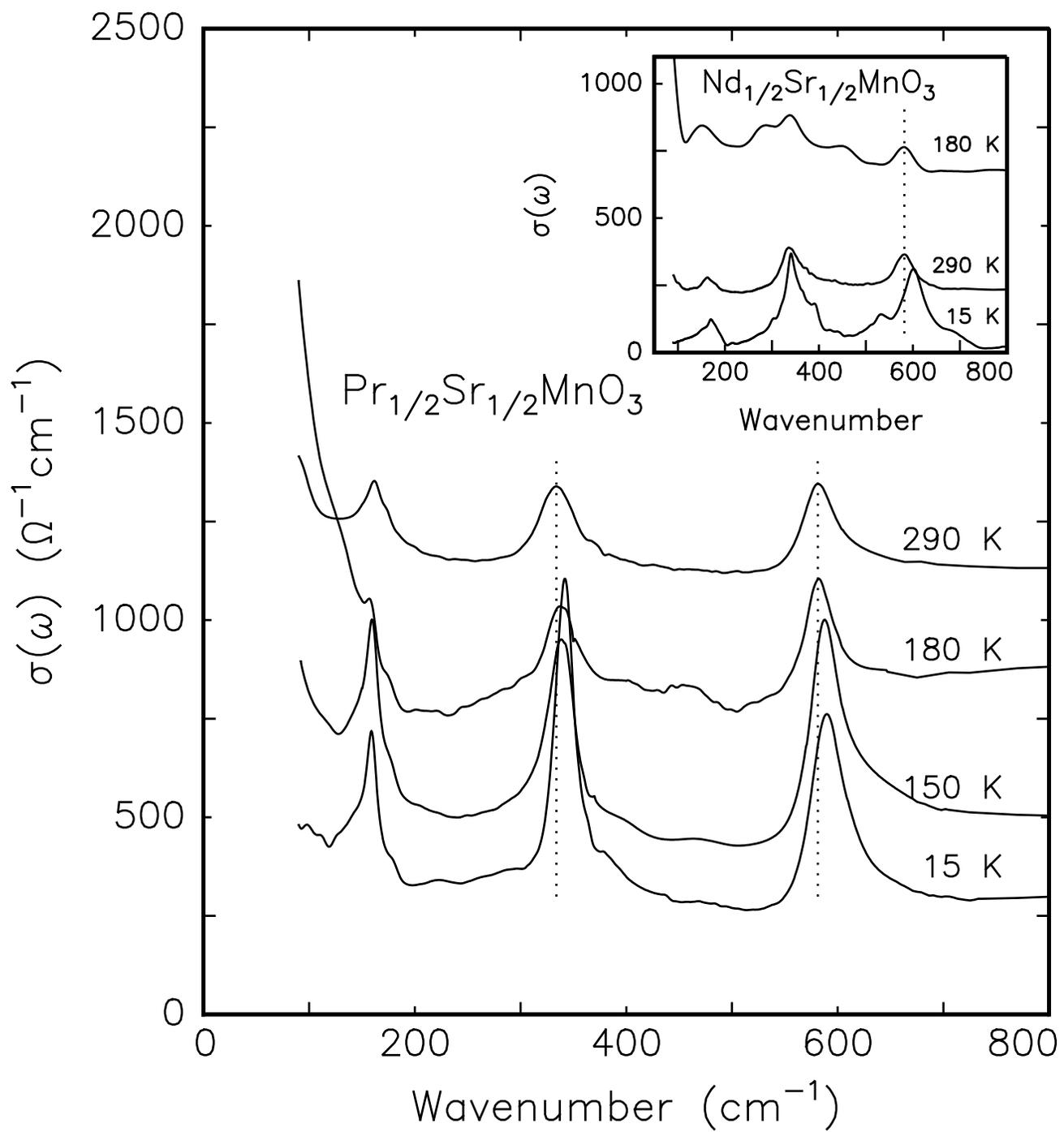
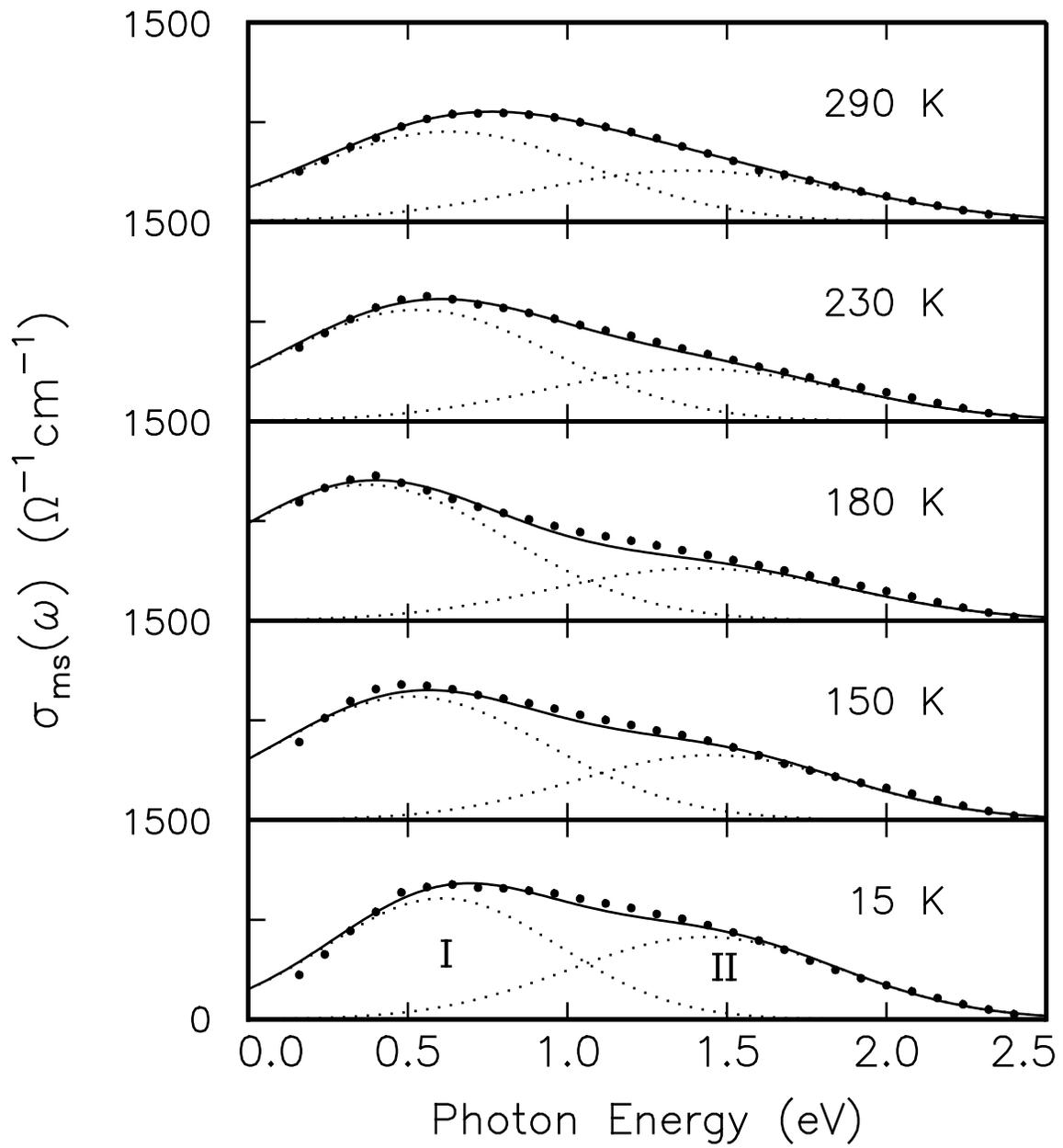
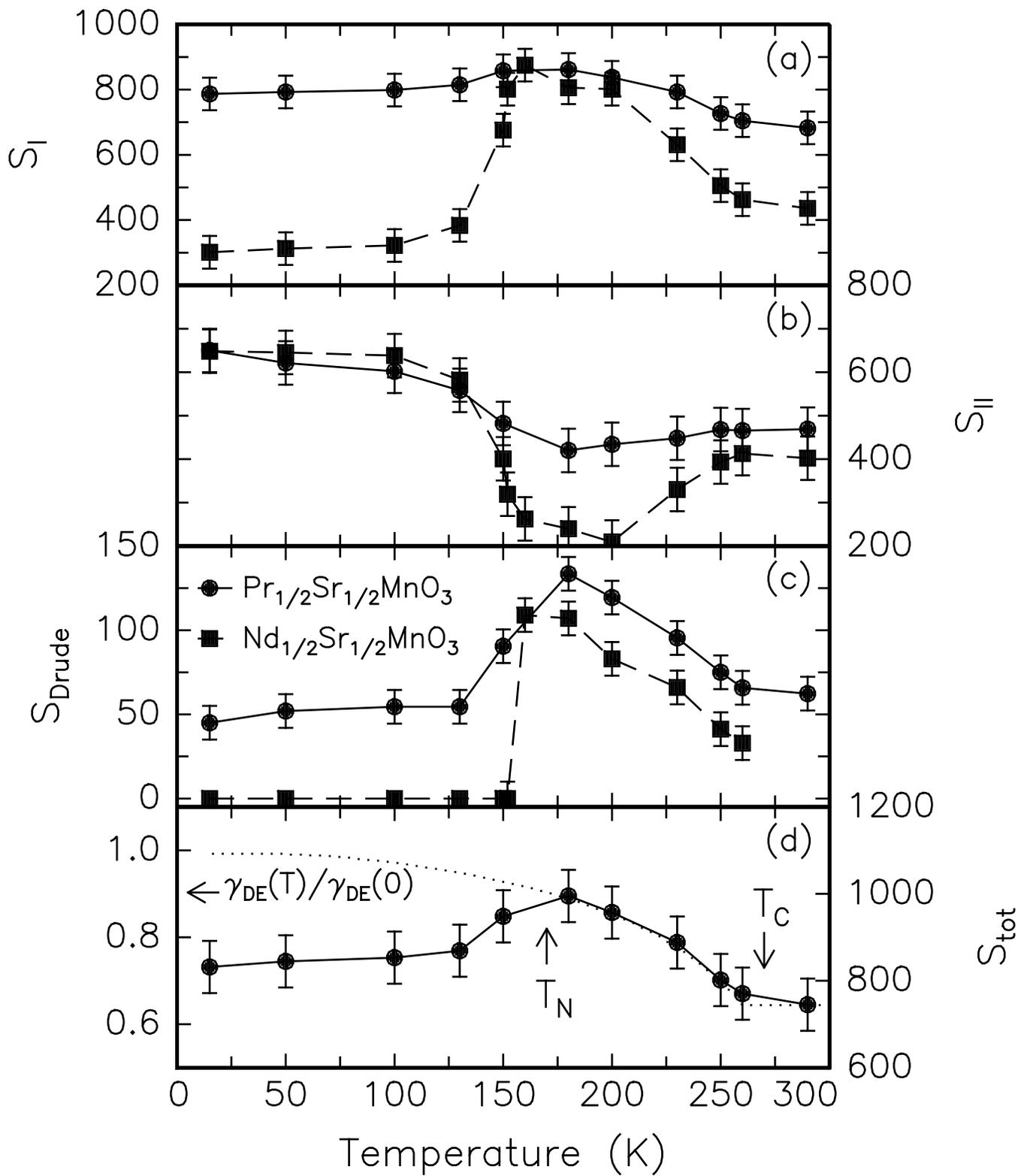
\end{document}